\documentclass[aps,prl,floatfix,showpacs,twocolumn]{revtex4}
\usepackage{amsmath,amssymb}
\usepackage{epsfig}

\newcommand{\bm}{\boldsymbol}

\begin{document}

\hsize\textwidth\columnwidth\hsize\csname@twocolumnfalse\endcsname

\title{Spin Hall Effect in Doped Semiconductor Structures}

\author{Wang-Kong Tse} 
\email[Electronic address:$\;$]{wktse@umd.edu}
\author{S. Das Sarma}
\affiliation{Condensed Matter Theory Center, Department of Physics,
University of Maryland, College Park, Maryland 20742}

\begin{abstract}
In this Letter we present a microscopic theory of the extrinsic spin
Hall effect based on the diagrammatic perturbation theory. Side-jump (SJ) and
skew-scattering (SS) contributions are explicitly taken into account to calculate
the spin Hall conductivity, and we show their effects scale as $\sigma_{xy}^{SJ}
/\sigma_{xy}^{SS}\sim(\hbar/\tau)/\varepsilon_F$, with $\tau$ being the transport relaxation time.
Motivated by recent experimental work we apply our theory to n- and p-doped $3$D and $2$D GaAs structures, obtaining $\sigma_s/\sigma_c \sim 10^{-3}-10^{-4}$ where $\sigma_{s(c)}$ is the spin Hall (charge) conductivity, which is in reasonable agreement with the recent experimental results of Kato {\textit{et al.}} [Science \textbf{306}, 1910 (2004)] in n-doped $3$D GaAs system.
\end{abstract}
\pacs{73.43.-f; 72.25.Dc; 75.80.+q; 71.70.Ej}

\maketitle
\newpage

Spin Hall effect (SHE) is an intriguing phenomenon, theoretically predicted \cite{Dyak1,Dyak2} in 1971, where the application of a longitudinal electric field creates a transverse motion of spins, with the spin-up and spin-down carriers transversing in perpendicular directions with respect to the electric field opposite to each other leading to a transverse spin current and presumably to spin accumulation at the edges of a bulk sample. There have been enormous recent interest and activity in this topic due to a number of reasons: (1) The emergence of the subject of ``spintronics'' \cite{RMP} where active control and manipulation of spin dynamics in electronic materials leads to conceptually new functionalities and projected novel device applications; (2) the re-discovery \cite{EH} of the original prediction of SHE, and the theoretical prediction and controversy \cite{IH,IHde,PhysT} surrounding a new type of SHE, called the ``intrinsic'' SHE \cite{IH}; and (3) recent reports of the experimental observation \cite{Aws,Aws2,wunder} of SHE in n- and p-doped semiconductor structures by two experimental groups. The fact that the two experimental groups report SHE observations differing by orders of magnitude in strength coupled with claims and counter-claims on whether the experimentally observed effects are the original (in this context referred to as the ``extrinsic'') SHE \cite{Dyak1,Dyak2} or the new intrinsic \cite{IH} SHE have made the subject of SHE one of the most intensively studied current topics in electronic solid state physics. The theory of SHE is in a flux -- although the original extrinsic SHE (arising from the spin-orbit coupling effect in impurity scattering) is on fairly firm conceptual ground, its magnitude has often been claimed to be minuscule (and far too weak to be of any experimental consequence) whereas the very existence of the intrinsic SHE (which arises from the intrinsic spin-orbit coupling effects in the band structure) has often been questioned. Therefore, taking the most pessimistic (optimistic) view of the theoretical literature an impartial observer could reasonably conclude that the SHE is essentially zero (very large in magnitude) always (often)! 

It is therefore quite important to provide, not just general theoretical frameworks, but concrete theoretical calculations relating to the specific experimental SHE measurements. In this Letter we provide one such concrete calculation \cite{rem} for the recent experimental measurements \cite{Aws,Aws2,wunder} using entirely the extrinsic SHE perspective, completely ignoring any intrinsic SHE considerations \cite{InZhan}. The fact that we get reasonable qualitative agreement with the experimental SHE data from one group \cite{Aws,Aws2}, but not the other \cite{wunder}, is significant in the context of the continuing debate and controversy in the SHE. In particular, our results support the claim \cite{Aws,Aws2} that the observed SHE in n-doped GaAs is the extrinsic, and not the intrinsic SHE. We also emphasize that the extrinsic SHE is always present in a doped semiconductor structure (since it arises from impurity scattering) although its magnitude can be small \cite{InZhan} depending on the strength of the spin-orbit coupling.

In this Letter we confine ourselves to the
extrinsic spin Hall effect, which arises from the effects of impurity scattering\cite{Dyak1,Dyak2}.
Drawing similarity with the well-studied anomalous
Hall effect \cite{Bruno1}, we calculate the spin Hall
conductivity using diagrammatic perturbation theory. In
particular, we derive general expressions for the side-jump and the
skew-scattering contributions using the Kubo-Greenwood formula, and
then apply the theory, within simplified model approximations, to $3$D and $2$D doped semiconductor systems obtaining simple analytical formulas for the magnitude of the extrinsic SHE. In $2$D we find the spin Hall conductivity to be formally equivalent to the corresponding anomalous Hall conductivity.

The single-particle Hamiltonian in the presence of spin-orbit (SO)
scattering due to impurities is
\begin{eqnarray}
H &=& H_0+H_{SO}+V \nonumber \\
&=&\frac{\left\vert\boldsymbol{p}-e\boldsymbol{A}/c\right\vert^2}{2m}+\frac{\lambda_0^2}{4}
\left[\boldsymbol{\sigma}\times\boldsymbol{\nabla}V(\boldsymbol{r})\right]\cdot\boldsymbol{p}+V(\boldsymbol{r}),
\label{eq1}
\end{eqnarray}
where $m$ is the carrier effective mass, $\lambda_0$ is a length characterizing
 the strength of SO interaction, $\bm{A}$ is the
vector potential, all other notations are
standard. We note that the SO coupling strength parameter $\lambda_0$ is greatly enhanced in the solid state
GaAs environment over its free-electron vacuum value of $\hbar/m_e c$, the Compton wavelength. This could be construed as a band-structure-induced renormalization of the effective Compton wavelength (just as the effective mass and the background lattice dielectric constant are modified) in the semiconductor environment. In GaAs this renormalization of the Compton length over its vacuum value is by as much as a factor of $10^3$ (or larger) reflecting the much stronger (by $10^6$) SO coupling in semiconductors over the corresponding vacuum effect. This leads to a much enhanced extrinsic SHE in GaAs than the corresponding vacuum estimate. The inclusion of this band structure effect in the SO coupling is the key to understanding the extrinsic SHE in recent experiments \cite{Aws,Aws2}, as has also been emphasized recently by Engel \textit{et al}. \cite{rem} in the context of a Boltzmann theory calculation.

We model the impurity scattering potential $V(\bm{r})$ as a short-range white noise disorder
with
$\langle V(\bm{r}_1)V(\bm{r}_2) \rangle =
n_iv_0^2\;\delta(\bm{r}_1-\bm{r}_2)$ and $\langle
V(\bm{r}_1)V(\bm{r}_2)V(\bm{r}_3)\rangle =
n_iv_0^3\;\delta(\bm{r}_1-\bm{r}_2)\;\delta(\bm{r}_2-\bm{r}_3)$, here
$n_i$ is the impurity density and $v_0$ is the Fourier component of
$V(\bm{r})$ at $q = 0$, which is related to the scattering amplitude
$f(\theta)$ as:
\begin{equation}
v_0 = \int \mathrm{d}^3 r\; V(\bm{r}) = -\frac{4\pi
\hbar^2}{m}f(\theta = 0) \label{eq2}
\end{equation}
The extrinsic spin Hall effect results from the SO coupling in two
ways: the antisymmetry of the matrix element $\langle \bm{k}\vert
H_{SO}\vert \bm{k}'\rangle$ with respect to interchanging $\bm{k}$ and
$\bm{k}'$ gives rise to the skew scattering while the
non-commutativity of $\boldsymbol{r}$ and $H_{SO}$, which results in
an extra term -- the anomalous current, gives rise to the side jump leading 
to a renormalization of the current vertex. This is similar 
to the anomalous Hall effect (AHE) since the same physics
underlies both AHE and SHE.

In the following we proceed to calculate the vertex correction of the spin current (i.e. the side-jump
contribution). The velocity is given by
\begin{eqnarray}
\boldsymbol{u} &=& \frac{i}{\hbar}\left[H,\boldsymbol{r}\right] \nonumber \\
&=&\frac{1}{m}\left({\boldsymbol{p}-\frac{e}{c}\bm{A}}\right)+\frac{\lambda_0^2}{4}
\left[\boldsymbol{\sigma}\times\boldsymbol{\nabla}V(\boldsymbol{r})\right],\label{eq3}
\end{eqnarray}
from which the spin current is calculated as
\begin{eqnarray}
\boldsymbol{j}_{s} &=& \frac{e}{4}\{\sigma_z,\boldsymbol{u}\} \nonumber \\
&=&e\left[\frac{1}{2m}\left({\boldsymbol{p}-\frac{e}{c}\bm{A}}\right)\sigma_z+\frac{\lambda_0^2}{8}\boldsymbol{\hat{z}}
\times\boldsymbol{\nabla}V(\boldsymbol{r})\right], \label{eq4}
\end{eqnarray}
here, following standard practice, we have multiplied the conventional definition of the spin
current by the electronic charge $e$ so that it has the same units and dimensions as the
charge current. Notice that the second term in Eq.~(\ref{eq4}) is
the `anomalous' contribution to the spin current.

\begin{figure}
  \includegraphics[width=8.5cm,angle=0]{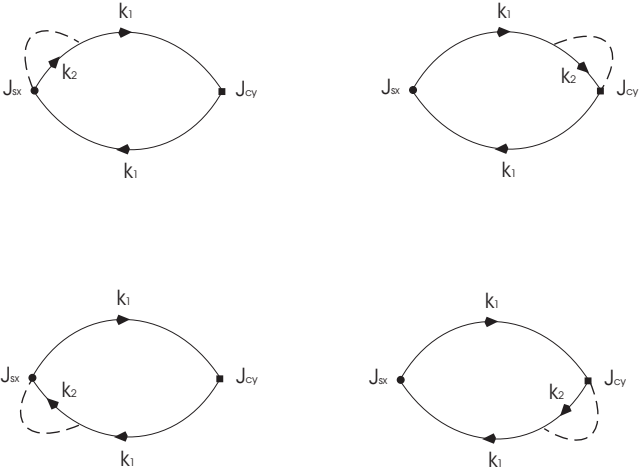}
\caption{Diagrams for the side-jump contribution, the vertex
correction for the spin current Eq.~(\ref{eq8}) is denoted by a
filled circle connected to a dashed line on the left-side vertex;
and the vertex correction for the charge current Eq.~(\ref{eq9}) by
a filled square connected to a dashed line on the right-side vertex.
The filled circle or square without connecting to a dashed line
implies a bare vertex.} \label{fig1}
\end{figure}

Expressing all quantities from now on in the momentum representation the 
Hamiltonian in the second quantized form can be written as (we use the same notation
$H$ to express the Hamiltonian in the first and the second quantized notations):
\begin{eqnarray}
H &=& \int \mathrm{d}^3 r\; \psi^{\dag}(\bm{r})\;H\;\psi(\bm{r}) \nonumber \\
&=& \sum_{\bm{k}\bm{k}'} \psi^{\dag}(\bm{k})\bigg\{\frac{1}{2m}\left\vert\hbar \bm{k}-\frac{e}{c}\bm{A}\right\vert^2\delta_{\bm{k}\bm{k}'} \nonumber \\
&&+V(\boldsymbol{k}-\boldsymbol{k}')\bigg[1+\frac{i\lambda_0^2}{4}(\bm{k}\times\bm{k}')\cdot\bm{\sigma})\bigg]\bigg\}\psi(\bm{k}'),
\label{eq5}
\end{eqnarray}
from which the SO vertex correction to the Green's function is identified as
\begin{equation}
\langle \bm{k}\vert \delta H_{SO} \vert \bm{k}' \rangle =
\frac{i\lambda_0^2}{4}(\bm{k}\times\bm{k}')\cdot\bm{\sigma},
\label{eq6}
\end{equation}
The spin current, in second quantized form, is
\begin{eqnarray}
\boldsymbol{J}_s &=& \int \mathrm{d}^3 r\; \psi^{\dag}(\boldsymbol{r})\;\bm{j}_s\; \psi(\bm{r}) \nonumber \\
&=&e\sum_{\bm{k}\bm{k}'}\psi^{\dag}(\bm{k})\bigg\{\frac{\hbar \boldsymbol{k}}{2m} \sigma_z
\delta_{\bm{k}\bm{k}'} \nonumber \\
&&+\frac{i\lambda_0^2}{8}V({\boldsymbol{k}-\boldsymbol{k}'})
\left[\bm{\hat{z}}\times\left(\bm{k}-\bm{k}'\right)\right]\bigg\}
\psi(\bm{k}'), \label{eq7}
\end{eqnarray}
from which the SO vertex renormalization for the spin current can be
identified as
\begin{equation}
\delta j_{s,l} = \frac{ie\lambda_0^2}{8}
\epsilon_{lmz}(k'_m-k_m)V(\bm{k}-\bm{k}'). \label{eq8}
\end{equation}
The charge current vertex renormalization can also be obtained as:
\begin{equation}
\delta j_{c,l} = \frac{ie\lambda_0^2}{4}
\epsilon_{lmn}(k'_m-k_m)\sigma_n V(\bm{k}-\bm{k}'). \label{eq9}
\end{equation}
From the Kubo-Greenwood formula, the spin Hall conductivity can be calculated
from the spin current-charge current correlation function, which
can be represented schematically as a bubble diagram with one spin current vertex and one 
charge current vertex (Fig.~\ref{fig1}).

\begin{figure}
  \includegraphics[width=8.5cm,angle=0]{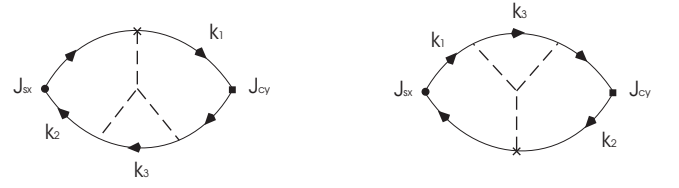}
\caption{Diagrams for the skew scattering contribution, the
correction to the Green's function Eq.~(\ref{eq6}) is denoted by a
cross.} \label{fig2}
\end{figure}

The retarded and advanced bare Green's function (i.e. without the SO
scattering potential) at zero frequency are diagonal matrices with
the diagonal elements
\begin{equation}
G^{R,A}_{\uparrow\uparrow,\downarrow\downarrow}(\bm{k}) =
\left(\varepsilon_{F\uparrow,\downarrow}-\varepsilon_{\uparrow,\downarrow}(k)\pm
i\hbar/2\tau_{\uparrow,\downarrow}\right)^{-1}. \label{eq10}
\end{equation}
where $\uparrow,\downarrow$ signifies spin-up and spin-down species,
$\varepsilon_{F\uparrow,\downarrow}$ their Fermi levels,
$\varepsilon_{\uparrow,\downarrow} = \hbar^2
k_{\uparrow,\downarrow}^2/2m$ the corresponding kinetic energies.
$\tau_{\uparrow,\downarrow}$ are the relaxation times for spin-up and
spin-down carriers in the first-order Born approximation given analytically for our
short-range scattering as $\tau_{\uparrow,\downarrow} = {\hbar}/{2\pi
N_{\uparrow,\downarrow} n_iv_0^2}$, with $N_{\uparrow,\downarrow}$ the density of states of the spin-up
and spin-down carriers at their respective Fermi levels.
Now taking into account of the vertex renormalizations of both spin
and charge currents (Fig.\ref{fig1}), we find the side-jump
contribution as:
\begin{eqnarray}
&&\sigma_{xy}^{SJ} = -\frac{ie^2}{8m}\;\lambda_0^2\; n_i v_0^2\;
\mathrm{tr}\sum_{\bm{k}_1\bm{k}_2} \nonumber \\
&&\bigg\{k_{1y}^2G^R(\bm{k}_1)G^A(\bm{k}_1)
\left[G^R(\bm{k}_2)-G^A(\bm{k}_2)\right]\bigg\}. \label{eq12}
\end{eqnarray}
The skew-scattering contribution can be obtained by keeping up to
third order in the scattering potential (Fig.~\ref{fig2}). Summing
both diagrams in Fig.~\ref{fig2}, we obtain:
\begin{eqnarray}
\sigma_{xy}^{SS} &=&\frac{ie^2\hbar^2}{16\pi m^2}\;\lambda_0^2\; n_i v_0^3\;\mathrm{tr}\sum_{\bm{k}_1\bm{k}_2\bm{k}_3} \nonumber \\
&&\bigg\{k_{1x}^2
G^R(\bm{k}_1)G^A(\bm{k}_1) k_{2y}^2
G^R(\bm{k}_2)G^A(\bm{k}_2) \nonumber \\
&&\left[G^R(\bm{k}_3)- G^A(\bm{k}_3)\right]\bigg\}. \label{eq13}
\end{eqnarray}
%
%
Evaluating the integral in Eqs.~(\ref{eq12})-(\ref{eq13}) for the
cases of $3$D and $2$D gives
\begin{eqnarray}
\sigma_{xy}^{SJ} = \left\{ \begin{array}{l}
-{e^2 \hbar}/{12m} \\
-{e^2 \hbar}/{8m}
\end{array} \right\}\lambda_0^2 \left(k_{F\uparrow}^2 N_{\uparrow}^2+k_{F\downarrow}^2 N_{\downarrow}^2 \right).
\begin{array}{l}
\;(3 \textrm{D}) \\
\;(2 \textrm{D})
\end{array}
\label{eq14}
\end{eqnarray}
\begin{eqnarray}
\sigma_{xy}^{SS} &=& \left\{ \begin{array}{l}
-{\pi e^2 \hbar^2}/{36 m^2} \\
-{\pi e^2 \hbar^2}/{16 m^2}
\end{array} \right\} \nonumber \\
&&\lambda_0^2 v_0 \left\{k_{F\uparrow}^4 N_{\uparrow}^2 \tau_{\uparrow}+k_{F\downarrow}^4 N_{\downarrow}^2
\tau_{\downarrow}\right\}.
\begin{array}{l}
\;\;\;\;\;(3 \mathrm{D}) \\
\;\;\;\;\;(2 \mathrm{D})
\end{array}
\label{eq15}
\end{eqnarray}
In spin Hall effect, one observes the deflection of opposite spins
from an unpolarized electron beam, so we set $n_{\uparrow} =
n_{\downarrow} = n/2$. Here the densities of states per spin
for 3D and 2D are
\begin{eqnarray}
N_{\uparrow,\downarrow} = \left\{ \begin{array}{ll}
{mk_F}/{2\pi^2\hbar^2}, &\;\;\;\;\;\;\;\;\;\;(3 \textrm{D}) \\
{m}/{2\pi\hbar^2}. &\;\;\;\;\;\;\;\;\;\;(2 \textrm{D})
\end{array} \right.
\label{eq16}
\end{eqnarray}
Eqs.~(\ref{eq14})-(\ref{eq16}) are our important extrinsic SHE results for $2$D and $3$D semiconductor 
structures within the short-range impurity scattering model. We believe that these analytic formulas would approximately apply even when the impurity scattering is not strictly zero-range, e.g. for screened ionized scattering. Assuming that the dominant impurities in the semiconductor to be randomly distributed screened ionized impurity centers, we can further simplify our SHE formulas for $2$D and $3$D doped semiconductor systems. In the first-order Born approximation, the scattering amplitude of
the screened Coulomb potential (assuming Thomas-Fermi screening) is
$f(\theta) = ({2me^2}/{\hbar^2 \epsilon_m})(q^2+q_{\mathrm{TF}}^2)^{-1}$ in $3$D; and 
$f(\theta) = ({me^2}/{\hbar^2 \epsilon_m})(q+q_{\mathrm{TF}})^{-1}$ in $2$D, 
where $\epsilon_m$ is the dielectric constant of the material, $q =
2k\,\textrm{sin}(\theta/2)$ is the momentum transfer. The Thomas-Fermi screening wavenumber is given by $q_{\mathrm{TF}} = \sqrt{6\pi ne^2/\epsilon_m \varepsilon_F}$ for $3$D and $q_{\mathrm{TF}} = 2\pi ne^2/\epsilon_m \varepsilon_F$ for $2$D. Using these formulas for the cases of $2$D and $3$D, it is then straightforward to calculate the extrinsic SHE contributions using Eq.~(\ref{eq14})-(\ref{eq16}) within this simplifying approximation scheme. Accordingly the side-jump contribution is found to be
\begin{equation}
\sigma_{xy}^{SJ} = -\frac{e^2\lambda_0^2}{4\hbar}n, \label{eq18}
\end{equation}
for both $3$D and $2$D cases, whereas for skew scattering we find
\begin{eqnarray}
\sigma_{xy}^{SS} = \left\{ \begin{array}{l}
{\pi m \lambda_0^2 \varepsilon_F}/{3\hbar^2} \\
{\pi m \lambda_0^2 \varepsilon_F}/{2\hbar^2}
\end{array} \right\}
\frac{ne^2\tau}{m}.
\begin{array}{l}
\;\;\;\;\;\;\;\;\;\; (3\textrm{D}) \\
\;\;\;\;\;\;\;\;\;\; (2\textrm{D})
\end{array}
\label{eq19}
\end{eqnarray}
We note that the side-jump and the skew-scattering contributions scale as $\sigma_{xy}^{SJ}/\sigma_{xy}^{SS} \sim (\hbar/\tau)/\varepsilon_{F}$. Typically $\tau \sim 10^{-13}-10^{-12}\,\mathrm{s}$, so the side-jump contribution is approximately comparable to the skew-scattering contribution when the Fermi energy $\varepsilon_F \sim 1-10\,\mathrm{meV}$.

Some order-of-magnitude estimates from
Eqs.~(\ref{eq18})-(\ref{eq19}) are in order. We choose $\lambda_0 = 4.7\times10^{-8}\,\mathrm{cm}$ (which is a factor of $10^3$ enhancement over the vacuum electron Compton wavelength of $3.9\times10^{-11}\,\mathrm{cm}$) in n-GaAs consistent with the expected electronic spin-orbit coupling strength in GaAs \cite{incc}, which is thus enhanced by six orders of magnitude over the corresponding Thomas term in the free electron vacuum case. For $3$D, we employ the
parameters from the experiment of Kato \textit{et al}. \cite{Aws,Aws2}, 
where the electron density
is $n = 3\times10^{16}\,\mathrm{cm}^{-3}$ and the longitudinal
conductivity $\sigma_{xx} \simeq 3\times10^{3}\,
\mathrm{\Omega}^{-1}\,\mathrm{m}^{-1}$. For $2$D, we take $n =
10^{11}\,\mathrm{cm}^{-2}$ and $\sigma_{xx} = 10^{-4}\,
\mathrm{\Omega}^{-1}$. We then get for $3$D $\sigma_{xy}^{SJ} =
-0.375\, \mathrm{\Omega}^{-1}\,\mathrm{m}^{-1}$ and
$\sigma_{xy}^{SS} = 2.97\, \mathrm{\Omega}^{-1}\,\mathrm{m}^{-1}$;
whereas for $2$D $\sigma_{xy}^{SJ} = -1.25\times 10^{-8}\,
\mathrm{\Omega}^{-1}\,\mathrm{m}^{-1}$ and  $\sigma_{xy}^{SS} =
10^{-7}\, \mathrm{\Omega}^{-1}\,\mathrm{m}^{-1}$. This gives, for
$3$D and $2$D respectively, the spin Hall conductivity as
$2.6\,\mathrm{\Omega}^{-1}\,\mathrm{m}^{-1}$ and $8.8\times
10^{-8}\,\mathrm{\Omega}^{-1}$; and the ratios of the spin Hall
conductivity to the longitudinal conductivity as $8.65\times
10^{-4}$ and $8.75\times 10^{-4}$. The experimental SHE quoted in Ref.~\cite{Aws}
is about $0.7 \, \mathrm{\Omega}^{-1}\,\mathrm{m}^{-1}$, which is a factor of $\sim 4$ smaller than our 
estimate. At this point we cannot ascertain whether this discrepancy is due to our highly simplified model or
the higher order corrections we have neglected here (e.g. vertex corrections 
due to diffuson pole or weak localization)
or the possible inaccuracies introduced in the experiment, where
 the spin Hall conductivity was not a
directly measured quantity but instead obtained by fitting the
measured spin accumulation data with a simple assumed spin density
profile. We also remark that the spin
accumulation data in \cite{Aws} was obtained in the presence of an
external magnetic field ($\sim 500\,\mathrm{mT}$),
which is known to decrease the spin accumulation due to spin
precession \cite{Dyak2, Johnson}. In addition, we note the crucial point that our 
calculated SHE is directly proportional to the SO coupling strength $\lambda_0^2$, 
which is only approximately known in GaAs -- any inaccuracy in the knowledge \cite{Winkler} of the precise 
SO coupling is directly reflected in our estimate of the extrinsic SHE. We also point out that the 
net SHE in our theory is a sum of a positive (skew-scattering) and a negative (side-jump) contribution, 
further leading to the possibility of quantitative errors.

Now we briefly extend our discussion to the case of $2$D holes using
the experimental parameters from the experiment of Wunderlich
{\textit{et al}}. \cite{wunder}, where the hole density is $p =
2\times10^{12}\,\mathrm{cm}^{-2}$ and longitudinal conductivity $\sigma_{xx} \simeq
1.09\times10^{-3}\mathrm{\Omega}^{-1}$. We obtain, assuming the same
spin-orbit coupling strength $\lambda_0$ as in the electron case, $\sigma_{xy}^{SJ}
= -1.55\times 10^{-8}\, \mathrm{\Omega}^{-1}$ and $\sigma_{xy}^{SS}
= 1.36\times 10^{-6}\, \mathrm{\Omega}^{-1}$, giving the total spin
Hall conductivity $1.34\times10^{-6}\, \mathrm{\Omega}^{-1}$ and
its ratio to the longitudinal conductivity as $1.23\times 10^{-3}$.
This is an order of magnitude lower than the intrinsic value estimated 
in Ref.~\cite{wunder}. We note, however, that the SO coupling strength parameter
for holes is expected to be larger than that for electrons, and therefore the possibility (at least
as a matter of principle) that the $2$D hole experiment in Ref.~\cite{wunder} is also a measurement 
of the extrinsic SHE cannot be completely ruled out.

We finally discuss the formal connection between the spin Hall effect
and the anomalous Hall effect. First the
side-jump contribution of the anomalous Hall conductivity
$\tilde{\sigma}_{xy}$ for both $3$D and $2$D can formally be expressed as
$\tilde{\sigma}_{xy}^{SJ} = -({e^2\lambda_0^2}/{2\hbar})s$, 
where $s = n_{\uparrow}-n_{\downarrow}$ is the spin density. Except
for the appearance of $s$ instead of $n$, this formula is very similar to 
that for the spin Hall effect, Eq.~(\ref{eq18}). The skew scattering contribution 
can be written as:
\begin{eqnarray}
\tilde{\sigma}_{xy}^{SS} = {\pi \varepsilon_F}/{mc^2}
\left\{ \begin{array}{l}
2se^2\tau_{\mathrm{eff}}/3m, \\
se^2\tau/m.
\end{array} \right\}
\begin{array}{l}
\;\;\;\;\;\ (3\textrm{D}) \\
\;\;\;\;\;\ (2\textrm{D})
\end{array}
\label{eq21}
\end{eqnarray}
\newline
where we have defined $\tau_{\mathrm{eff}} =
(k_{F\uparrow}^6\tau_{\uparrow}-k_{F\downarrow}^6\tau_{\downarrow})/(k_{F\uparrow}^6-k_{F\downarrow}^6)$
for the case of $3$D to be an effective relaxation time. Now we note that both 
the skew-scattering and the side-jump contributions for the anomalous Hall effect
 and the spin Hall effect appear in very similar forms except for an extra factor of $2$ which comes from
$\sigma_z /2$ in the definition of the spin Hall current
Eq.~(\ref{eq4}). In the particular case of $2$D, the relaxation times
for spin-up and spin-down carriers are equal $\tau_{\uparrow} =
\tau_{\downarrow}$ since the density of states per spin is a
constant, we have the equality between the spin Hall \textit{mobility} and 
the anomalous Hall \textit{mobility} (except for a factor of two as explained above):
\begin{equation}
\frac{\sigma_{xy}}{n} = \frac{\tilde{\sigma}_{xy}}{2s}. \label{eq24}
\end{equation}
%

In summary, we have obtained analytic formulas for the extrinsic spin Hall effect in $2$D and $3$D doped 
semiconductor structures using a Kubo formula-based diagrammatic perturbation theory, our
theory explicitly manifests the formal connection between the SHE and the AHE. We find that the recent
 experimental results of Kato \textit{et al}. \cite{Aws} are consistent with our (admittedly crude) estimate 
of the extrinsic SHE whereas the experimental results of Wunderlich \textit{et al.} \cite{wunder}, while being larger 
than our estimate, are not impossibly large given the uncertainty in the precise knowledge of the SO 
coupling strength. We therefore suggest the possibility, at least as a matter of principle, that the recent 
experimental observations of the SHE in GaAs are the confirmation of a beautiful prediction 
\cite{Dyak1,Dyak2} going back more than thirty years.
 
This work is supported by NSF, ONR, and LPS-NSA.


\end{document}